\documentclass[%
 aip,
 jap,
 amsmath,amssymb,
 reprint,%
]{revtex4-1}

\usepackage{graphicx}
\usepackage{dcolumn}
\usepackage{bm}

\begin{document}

\title[Sample title]{Geometry induced entanglement transitions in nanostructures}

\author{J. P. Coe}
 \email{jpc503@york.ac.uk}
\author{S. Abdullah}%
 \email{sma503@york.ac.uk}
\author{I. D'Amico}
\email{ida500@york.ac.uk}
 \affiliation{Department of Physics, University of York, York YO10 5DD, UK}

\begin{abstract}
We model quantum dot nanostructures using a one-dimensional system of two interacting electrons.  We show that strong and rapid variations may be induced in the spatial entanglement by varying the nanostructure geometry.  We investigate the position-space information entropy as an indicator of the entanglement in this system.  We also consider the expectation value of the Coulomb interaction and the ratio of this expectation to the expectation of the confining potential and their link to the entanglement.  We look at the first derivative of the entanglement and the position-space information entropy to infer information about a possible quantum phase transition.
\end{abstract}


\maketitle

\section{\label{sec:Introduction} Introduction}
For a long time entanglement had  been thought of as a failure of quantum mechanics or at best a quantum curiosity, but it is now considered a physical resource which could feasibly allow quantum information processors to overcome some of the limitations of their classical analogues.   

Semiconductor quantum dots have been put forward as possible components of solid state quantum information processing devices\cite{DV,IDA1,IDA2}.  One of the possible advantages of the use of quantum dots is that the parameters of the system may be tuned, allowing the properties of semiconductor nanostructures to be tailored\cite{IDA1}.  
The seemingly inexorable  progress of technology appears to promise advanced engineering of quantum dot-based structures, thus leading to the fabrication of coupled and scalable quantum dot systems.  To use  quantum dot devices for quantum computing necessitates the ability to generate and manipulate entanglement within these structures.  
In quantum dot systems this entanglement could be controlled through externally applied  electro-magnetic fields\cite{Legel} or by varying the parameters of the nanostructure\cite{Abdullah}.  In this contribution, we show how changing the geometry of the confinement potential of single, core-shell and double quantum dot structures can influence the spatial entanglement\cite{Coe} between two electrons trapped within the nanostructure;  such a property could possibly be exploited to design nanostructures according to the level of entanglement needed for a specific application.  
We also investigate the use of the position-space information entropy as an indicator of the entanglement for this system.  Our results show a potential quantum phase transition induced by  varying the nanostructure confinement geometry. A quantum phase transition is characterized by a nonanalyticity in the ground state energy and could be detected by the nonanalyticity of the entanglement measure \cite{WU1} which we seem to observe.  The  transition we observe could be created experimentally by changing the polarity of a gate positioned over the core-to-barrier region within a gate-defined quantum dot.

\section{The model}
We model the quantum dot system with a one dimensional Hamiltonian describing two interacting electrons constrained by a two-center power-exponential potential
\begin{equation}
	H=\sum_{i=1}^{2}{\left[-\frac{1}{2}\frac{d^{2}}{dx_{i}^{2}}+V(x_{i})\right]}+U(x_{1},x_{2}),
\end{equation}
where we employ (effective) atomic  units.  Here we model the Coulomb repulsion via the contact interaction\cite{DELTA1,DELTA2,DELTA3} $U(x_{1},x_{2})=\delta(x_{1}-x_{2})$  while the two-center power-exponential potential\cite{ADAMOWSKI} $V(x)$  is given by
\begin{equation}
V(x)=-V_{0}\left\{\exp[-(|x+d|/R)^{p}]+\exp[-(|x-d|/R)^{p}]\right\}.
\label{eqn:2centrePE} 
\end{equation}
This potential allows us to access a range of nanostructure geometries: single dots, double dots and core-shell dots (well-within-a-well)\cite{Abdullah}. We may move continuously from one structure to the next by varying the potential range $R$. For fixed potential well depth $V_{0}$ and well center distance $2d$, the parameter $p$ allows us to control the `hardness' of the potential.  
This, for example, allows us to consider quantum dots with little or no intermixing between the different materials using a large $p$ value (hard potential, p=$200$) or, by using a smaller $p$ value, $p\lesssim 7$, smoother (`softer') potentials  corresponding to the experimental situation where there is substantial mixing at the interface between materials.  We set $p=200$ so as to have almost square wells and thereby allow strong and very fast transitions in the entanglement to be observed. 
We solve the system using `exact' diagonalization based on a single-particle harmonic oscillator basis set\cite{Abdullah}.
\section{Entanglement}
As the spin-part of the ground state wavefunction is  constant (in fact a singlet: a maximally entangled qubit state) we investigate the spatial entanglement\cite{Coe}.
We measure the entanglement using the linear entropy of the reduced density matrix 
\begin{equation}
\nonumber
L=Tr (\rho_{\text{red}}- \rho_{\text{red}}^2)=1-Tr \rho_{\text{red}}^2.
\end{equation}
Here the reduced density matrix is $\rho_{\text{red}}=Tr_{B}\left|\psi \right\rangle \left \langle \psi \right|$ and the subsystem $B$ corresponds to one of the electrons. 
It has been shown that $L$ may be used as a valid measure of entanglement when quantifying  the particle-particle entanglement.\cite{BUSCEMI07,Coe} Importantly $L$ is much easier to calculate than $S$, the von Neumann entropy of the reduced density matrix\cite{NIELSEN}, when a system has a very large number of degrees of freedom. This is the case for the spatial entanglement where the degrees of freedom are continuous. In this case the reduced density matrix is given by
\begin{equation}
\nonumber
\rho_{\text{red}}(x_{1},x_{2})=\int \Psi^{*}(x_{1},x_{3})\Psi(x_{2},x_{3})dx_{3},
\end{equation}
with $\Psi$ the many-body wave-function,
\begin{equation}
\nonumber
\rho^{2}_{\text{red}}(x_{1},x_{2})=\int \rho_{\text{red}}(x_{1},x_{3})\rho_{\text{red}}(x_{3},x_{2})dx_{3},
\end{equation}
and
\begin{equation}
\nonumber
Tr\rho^{2}_{\text{red}}=\int \rho^{2}_{\text{red}}(x,x)dx. 
\end{equation}
For a maximally entangled system with continuous degrees of freedom $L$ would be equal to unity.

We also compare $L$ with the position-space information entropy of the density
\begin{equation}
S_{n}=-\int n(x)\log n(x)dx,
\label{eqn:infentropy}
\end{equation} where $n(x)$ is the particle density.
This has been used to study the entanglement of the Moshinsky atom \cite{AMOVILLI04} and
may also be thought of as a zeroth-order approximation to $S$, where the off-diagonal terms of the reduced density matrix are set to zero. \cite{Coe}
\section{Results}
\begin{figure}
\includegraphics[width=0.4\textwidth]{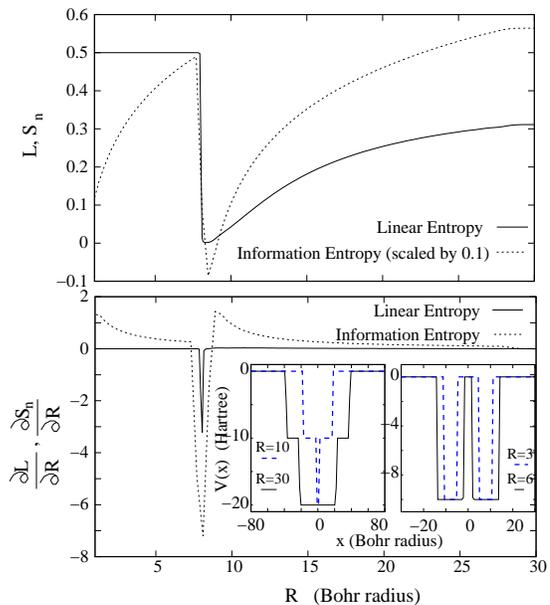}
\caption{\label{fig:Linear} Upper panel: The linear entropy of the reduced density matrix $L$ and the position-space information entropy of the density $S_{n}$ scaled by $0.1$.  Lower Panel: The partial derivative with respect to $R$ of $L$ and $S_{n}$. Inset: potential shape for different values of $R$, as labeled. }
\end{figure}

Our results show that the spatial entanglement can indeed be successfully manipulated through the geometry of the confining potential: in Fig.~\ref{fig:Linear}, upper panel, we show that, by manipulating the geometry of the nanostructure, we can create entanglement, as quantified by $L$, from almost zero to $0.5$ (the maximum for a system of two qubits).  For $R \sim 30$ the potential is a well-within-a-well for which the inner well becomes narrower with decreasing $R$ (see inset of Fig.~\ref{fig:Linear}) and the entanglement decreases until it is almost zero. In this case the electrons are now essentially described by a product state, the wave function dominated by the very strongly confining potential.  This inner well then inverts to become a barrier and gives a double well structure which has an entanglement of $\sim 0.5$. This entanglement is relatively robust: as $R$ decreases further, the distance and width of the double wells change (see inset of Fig.~\ref{fig:Linear}), but the entanglement does not.
The sharp  transition from almost zero entanglement to the maximal entanglement for a qubit system occurring at $R\approx 8$ in Fig.~\ref{fig:Linear}  means that we have moved from an essentially non-interacting non-entangled product state -- due to the very strongly confining inner potential well in the core-shell QD -- to a highly entangled state where the wave-function splits between the two wells.  We find that a discontinuity in the derivative (lower panel of Fig.~\ref{fig:Linear})  of the entanglement measure  with respect to the potential range $R$ seems to begin to appear at this point. This discontinuity  suggests a possible quantum phase transition.\cite{WU1}

When comparing $S_{n}$, the position-space information entropy (Eq.~\ref{eqn:infentropy}), to  the linear entropy $L$, we find that, for $R$ greater than the transition point,  $S_{n}$ has a behavior qualitatively similar to $L$. However for lower $R$ it does not display a constant value but instead sharply decreases. This confirms that $S_{n}$ cannot be considered as a measure, nor an indicator of entanglement.
It is interesting to notice though that the sharp entanglement transition  at $R\approx 8$ is captured by  the position-space information entropy $S_{n}$. 

\begin{figure}
\includegraphics[width=0.4\textwidth]{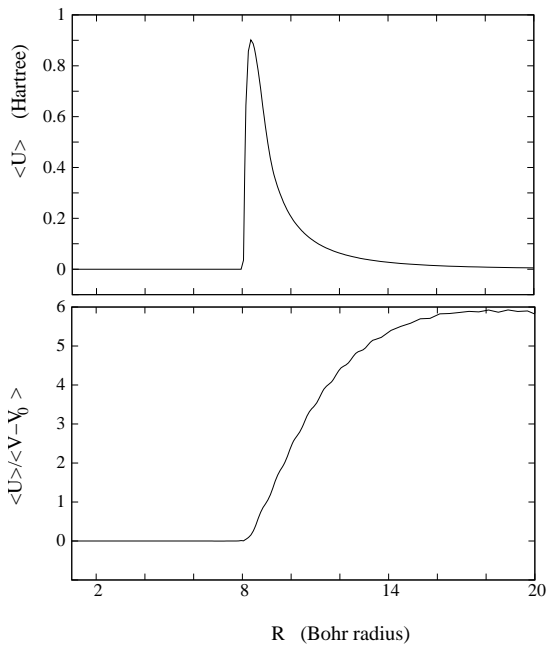}
\caption{\label{fig:Coul} Upper panel: The expectation value of the Coulomb interaction.  Lower panel:  The ratio of expectations of the Coulomb interaction and potential energy where the potential is shifted by its lowest value $V_{0}$.}
\end{figure}

Our results suggest that the entanglement has an inverse correlation with the Coulomb interaction energy for this system, shown in the upper panel of Fig.~\ref{fig:Coul}. 
The minimum of the interaction energy corresponds in fact to the maximum of the entanglement. 
This seems counter-intuitive as a non-zero spatial entanglement requires an interaction between the particles; however we note that while the Coulomb interaction strength is fixed, its expectation
value will be very low when the probability of finding the particles in the same region is almost zero.  
At the transition  ($R\approx 8$) the interaction energy falls from its maximum value -- where both electrons are confined in the inner well of a core-shell dot and there is almost zero entanglement-- to almost zero interaction energy, when the electrons are confined in a double well system and their repulsion means they are unlikely to be found in the same well.  

In previous work\cite{Coe} we suggested that for the three-dimensional Hooke's atom the ratio of the Coulomb interaction energy to the potential energy could be used to indicate the entanglement.  This was for a harmonic potential which always had zero as its lowest value.  Hence to allow a fair comparison we shift the potential of the system under investigation by its lowest value $V_{0}$, and plot $<U>/<V-V_0>$ in the lower panel of  Fig.~\ref{fig:Coul}.
 Due to the finite numerical precision in the difference $<V>-V_0$, we have to limit the data to $R\leq 20$.  
We see that, although $<U>/<V-V_0>$ indicates the trend of the entanglement for $R \gtrsim 8$ it does not pick up the large increase in entanglement for smaller $R$. We suggest this could be a consequence of the system topology and merits further investigation.  In addition this quantity does not experience a sharp change as the entanglement rapidly shifts from its minimum to its maximum, rather it appears to still be a smooth function in the range $R \approx 8$.
\section{Conclusions}
The results obtained for our model system results suggest that strong and rapid variations in the spatial entanglement in nanostructures may be induced by varying the potential geometry.  We found that the maximum entanglement we could reach was an entanglement of $L=0.5$ (the entanglement of two maximally-entangled qubits).  This occurred when the nanostructure was composed of a double well and this result was robust against the shape and distance of the two wells.  We have seen that the Coulomb interaction energy appears to indicate the entanglement through an inverse relationship; however neither the ratio of the interaction energy to potential energy nor the position-space information entropy could be considered as an entanglement indicator for this one-dimensional system. 
Our numerical results suggested that there might be a discontinuity in the first derivative of the entanglement with respect to $R$ for even harder confining potential ($p\to\infty$). This was suggestive of a potential quantum phase transition and corresponded to the transition between minimum and maximum entanglement.  Interestingly, a similar feature appeared also in  the first derivative of the  position-space information entropy with respect to $R$.

\end{document}